\documentclass[aps,twocolumn,showpacs]{revtex4}
\usepackage{graphicx}  % needed for figures
\usepackage{dcolumn}   % needed for some tables
\usepackage{bm}        % for math
\usepackage{amssymb}   % for math
\usepackage{amsmath}
\newcommand{\be}{\begin{eqnarray}}
\newcommand{\ee}{\end{eqnarray}}
\newcommand{\pp}[1]{\phantom{#1}}

\begin{document}

\title{Vacuum Rabi oscillation in nonzero-temperature open cavity}
\author{Patrycja Stefa\'nska,$^1$ Marcin Wilczewski,$^2$ and Marek Czachor$^2$}
\affiliation{
$^1$Zesp\'o{\l} Fizyki Atomowej --- Katedra Fizyki Atomowej i Luminescencji,
Politechnika Gda\'nska, 80-233 Gda\'nsk, Poland\\
$^2$Katedra Fizyki Teoretycznej i Informatyki Kwantowej, 
Politechnika Gda\'nska, 80-233 Gda\'nsk, Poland
}
\begin{abstract}
\begin{center}
\textbf{Published as: Open Syst.\ Inf.\ Dyn.\ 18 (2011) 363-373}\\
\textbf{DOI: 10.1142/S123016121100025X}
\end{center}
\mbox{} \\*[-3ex]
Comparison of theory of Rabi oscillations with experiment [M. Wilczewski and M. Czachor, Phys. Rev. A {\bf 79}, 033836 (2009)] suggests that cavity lifetime parameters obtained in measurements with many photons may be much smaller than those applicable to almost vacuum states of light. In this context we show that the conclusion remains unchanged even if one takes a more realistic description of the initial state of light in cavity.
\end{abstract}
\maketitle

\section{Introduction}

The 1996 Brune {\it et al.\/} experiment on vacuum Rabi oscillation \cite{Brune,Haroche} was analyzed in \cite{WC} by means of several alternative models of atom-reservoir interaction. The study was motivated by difficulties with fitting the data by theoretical curves, a problem addressed earlier by various authors \cite{ChoughPhD,Chough,Bonifacio}. Agreement with experimental Rabi oscillation data was then obtained but for the price of a cavity quality factor that was 500 times bigger than the one reported in \cite{Brune}. A part of open questions thus remained.

The solutions of master equations discussed in \cite{WC} were easier to find than in standard approaches because the formalism was based on jump operators generating transitions between the {\it dressed states\/} of the atom-field system. Such a construction is more consistent with the general theory of open systems \cite{Davies,Davies2} than the popular approach based on jumps between the atomic (hence bare) states \cite{Carmichael}, and is mathematically simpler. In the context of quantum optics it appeared only relatively recently in \cite{Scala}. Another reason why it was possible to find exact solutions was that the atom-field system was assumed to start from the initial photonic vacuum state. In terms of dressed states the initial state belonged to the subspace of the first dressed-state doublet.

The latter assumption was not very realistic. Light in the cavity was initially in thermal state at $T=0.8$K. The analysis from \cite{WC} not only took into account emissions from the dressed states downward on the energy ladder, but also thermal excitations from vacuum to the two dressed states, as well as thermal long-wave fluctuations between the dressed states from the first doublet. But it did not take into account the presence of the remaining bands of dressed states in the initial state. The first doublet appears with probability around $0.95$, but the second doublet has probability higher than $0.045$. It looks like downward transitions from the second doublet are processes of the same order as the upward thermal transitions at 0.8K from exact vacuum to the first doublet.
\begin{figure}
\includegraphics[width=5cm]{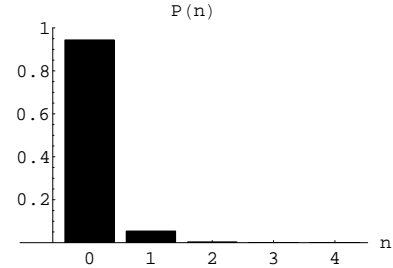}
\caption{Probabilities of vacuum ($n=0$) and 1-photon initial states ($n=1$) effectively dominate the initial thermal probability distribution at 0.8K. Contributions from $n\geq 1$ were not taken into account in \cite{WC}.}
\end{figure}
So what remained unclear in \cite{WC} was to what extent the fact that the initial state was thermal was influencing the behavior of the system. In order to understand the issue one could iteratively solve equations involving more and more doublets and more transitions between them, and compare predictions with the data. If inclusion of a next doublet would not produce visible modifications of Rabi oscillations, it would be justified to conclude that truncation of the Hilbert space to a subspace spanned by a given number of dressed state has sufficiently well approximated the thermal state.

The goal of the present paper is perform this test on the data from \cite{Brune}. We will see that inclusion of the next doublet of dressed states essentially complicates calculations, but does not really change agreement with experiment. We conclude that solution of the problem of the ``wrong" cavity $Q$ factor will not be achieved by taking more realistic initial states. The true physical mechanism must be therefore different.

\section{Open-cavity model}

We employ the standard Jaynes-Cummings Hamiltonian $H=\hbar \Omega$ in exact resonance,
\begin{eqnarray}
\Omega
&=&%1
\frac{\omega_0}{2}
\left( |e\rangle \langle e| -  |g\rangle \langle g|  \right)
\nonumber\\
&\pp=&
+
\omega_0
a^\dagger a
+
{\textsf{g}}
\left(
a |e\rangle \langle g|
+
a^\dagger
|g\rangle \langle e|
\right).
\label{eq:hjc}
\end{eqnarray}
The initial state at $T=0.8$~K is
\be
\rho(0)&=& p_0|e,0\rangle\langle e,0| + p_1|e,1\rangle\langle e,1|+\dots,
\ee
where $p_0=0.952381$, $p_1= 0.0453515$, $p_2=0.00215959$. Let us note that $\sum_{j=1}^\infty p_j=1-p_0=0.047619$ is of the order of $p_1$. The analysis performed in \cite{WC} assumed jumps between dressed states with transition coefficients typical of $T=0.8$~K, but the initial state was approximated by $\rho(0)= |e,0\rangle\langle e,0|$. The solution given in \cite{WC} employed transitions between the dressed states shown in Fig.~2, i.e. those involving $|e,0\rangle$ and the ground state:
\begin{eqnarray}
|\Omega_+\rangle
&=&
\frac {1}{\sqrt{2}}
\big(
|g,1\rangle + |e,0\rangle
\big),
\\
|\Omega_-\rangle
&=&
\frac {1}{\sqrt{2}}
\big(
|g,1\rangle - |e,0\rangle
\big),
\\
|\Omega_0\rangle
&=&
|g,0\rangle.
\end{eqnarray}
\begin{figure}
\includegraphics[width=5cm]{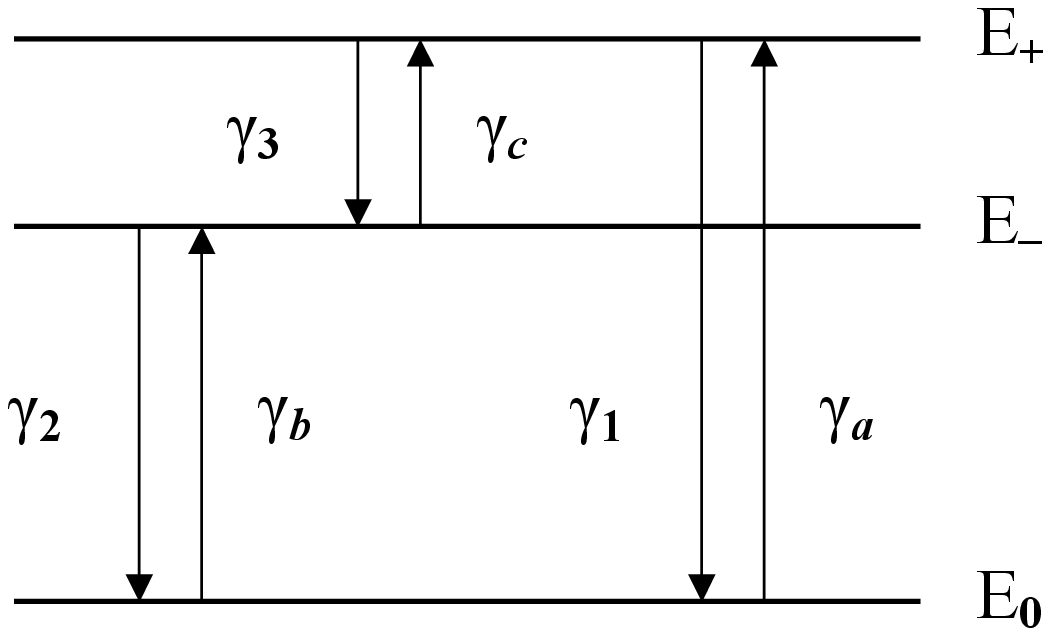}%{drabinka_stara1.eps}
\caption{Energy levels and decay coefficients used in the generalization of the Scala model \cite{Scala} discussed in \cite{WC}; $E_\pm=\hbar\Omega_\pm$, $E_0=\hbar\Omega_0$.}
\label{fig:drabinka}
\end{figure}
The decay coefficients from Fig.~2 satisfied
\be
\gamma_a
&=&%1
e^{-\frac{ \hbar (\omega_0 + {\tt{g}} ) }{ k T } }
\gamma_1
\approx
0.0466327
\gamma_1,
\\
\gamma_b
&=&%2
e^{-\frac{ \hbar (\omega_0 - {\tt{g}} ) }{ k T } }
\gamma_2
\approx
0.0466328
\gamma_2,
\\
\gamma_c
&=&%3
e^{-\frac{ 2 \hbar {\tt{g}}  }{ k T } }
\gamma_3
\approx
0.999997
\gamma_3,
\ee
Thermal excitations from the ground state to the first two dressed states involve proportionality factors
$e^{-\frac{ \hbar (\omega_0 - {\tt{g}})}{kT}}$ that are of the same order as $p_1$. Since $p_1$ measures the probability of occurrence of $|e,1\rangle\langle e,1|$ in the initial thermal mixture, it simultaneously determines probabilities of finding the next two dressed states. Accordingly, transitions determined by $\gamma_a$ and $\gamma_b$ should be regarded as processes of the same order as downward transitions from the second dressed-state doublet:
\begin{eqnarray}
|\Omega_{2+}\rangle
&=&
\frac {1}{\sqrt{2}}
\big(
|g,2\rangle + |e,1\rangle
\big),
\\
|\Omega_{2-}\rangle
&=&
\frac {1}{\sqrt{2}}
\big(
|g,2\rangle - |e,1\rangle
\big).
\end{eqnarray}
It is therefore justified to regard Fig.~3 as more realistic than Fig.~2.
The open-cavity generalization of the Scala et al. model \cite{Scala}, corresponding to Fig.~3, reads
\begin{figure}
		\includegraphics[width=5cm]{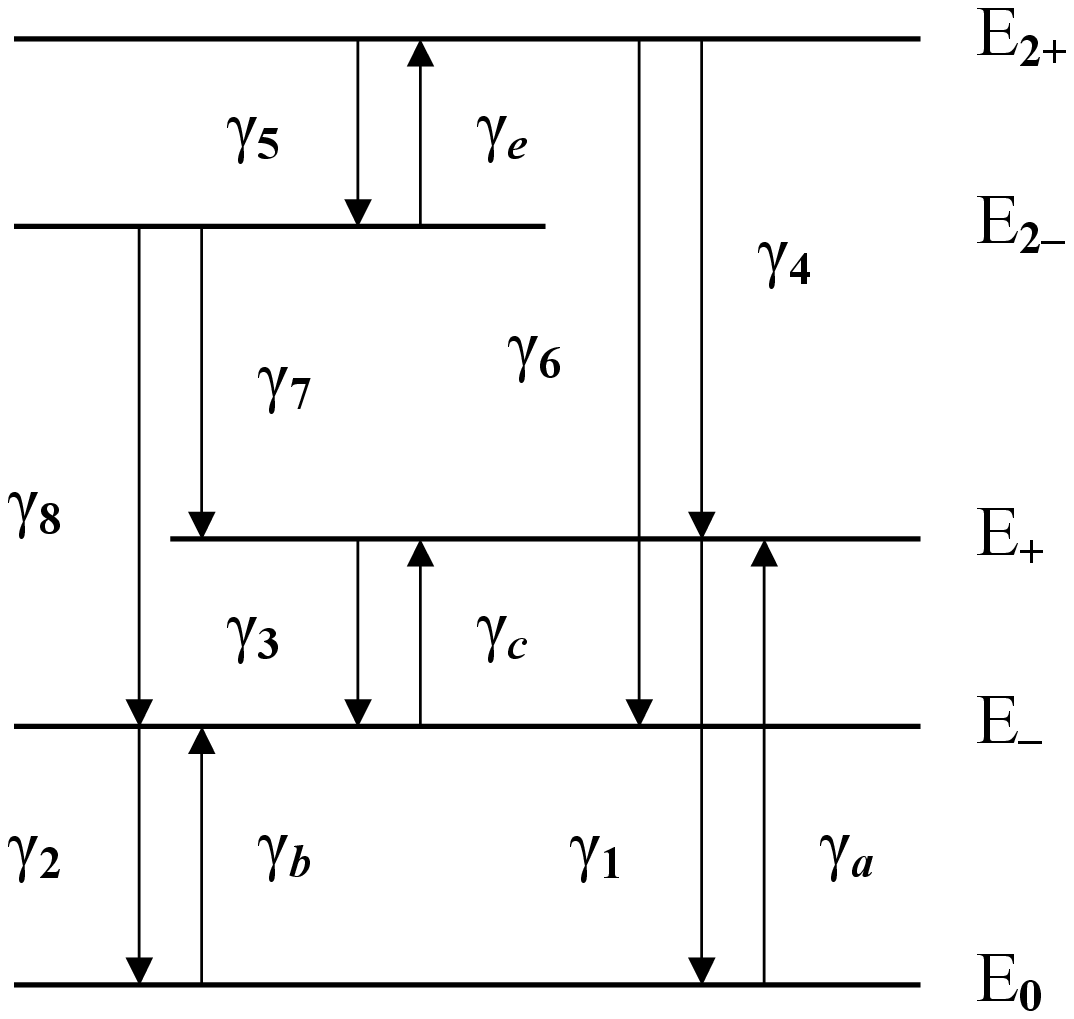}
	\caption{ $E_{2+}=\hbar\Omega_{2+}$,$E_{2-}=\hbar\Omega_{2-}$, $E_\pm=\hbar\Omega_\pm$, $E_0=\hbar\Omega_0$.}
	\label{fig:mój_schemat}
\end{figure}
\begin{widetext}
\begin{eqnarray}
\dot{\rho}
&=&
{\cal L}\rho
=
-i[\Omega,\rho]
\nonumber
\\
&\pp=&%6
+
\gamma_1
\left\{
\frac 1 2
|\Omega_0\rangle \langle \Omega_+| \rho |\Omega_+\rangle \langle \Omega_0|
-
\frac 1 4
\Big[| \Omega_+\rangle \langle \Omega_+|,\rho\Big]_+
\right\}
+
\gamma_a
\left\{
\frac 1 2
|\Omega_+\rangle \langle \Omega_0| \rho |\Omega_0\rangle \langle \Omega_+|
-
\frac 1 4
\Big[| \Omega_0\rangle \langle \Omega_0|,\rho\Big]_+
\right\}
\nonumber
\\
&\pp=&%6
+
\gamma_2
\left\{
\frac 1 2
|\Omega_0\rangle \langle \Omega_-| \rho |\Omega_-\rangle \langle \Omega_0|
-
\frac  1 4
\Big[| \Omega_-\rangle \langle \Omega_-|,\rho\Big]_+
\right\}
+
\gamma_b
\left\{
\frac 1 2
|\Omega_-\rangle \langle \Omega_0| \rho |\Omega_0\rangle \langle \Omega_-|
-
\frac  1 4
\Big[| \Omega_0\rangle \langle \Omega_0|,\rho\Big]_+
\right\}
\nonumber
\\
&\pp=&%6
+
\gamma_3
\left\{
\frac 1 2
|\Omega_-\rangle \langle \Omega_+| \rho |\Omega_+\rangle \langle \Omega_-|
-
\frac 1 4
\Big[| \Omega_+\rangle \langle \Omega_+|,\rho\Big]_+
\right\}
+
\gamma_c
\left\{
\frac 1 2
|\Omega_+\rangle \langle \Omega_-| \rho |\Omega_-\rangle \langle \Omega_+|
-
\frac 1 4
\Big[| \Omega_-\rangle \langle \Omega_-|,\rho\Big]_+
\right\}
\nonumber
\\
&\pp=&%6
+
\gamma_4
\left\{
\frac 1 2
|\Omega_+\rangle \langle \Omega_{2+}| \rho |\Omega_{2+}\rangle \langle \Omega_+|
-
\frac 1 4
\Big[| \Omega_{2+}\rangle \langle \Omega_{2+}|,\rho\Big]_+
\right\}
+
\gamma_5
\left\{
\frac 1 2
|\Omega_{2-}\rangle \langle \Omega_{2+}| \rho |\Omega_{2+}\rangle \langle \Omega_{2-}|
-
\frac 1 4
\Big[| \Omega_{2+}\rangle \langle \Omega_{2+}|,\rho\Big]_+
\right\}
\nonumber
\\
&\pp=&%6
+
\gamma_e
\left\{
\frac 1 2
|\Omega_{2+}\rangle \langle \Omega_{2-}| \rho |\Omega_{2-}\rangle \langle \Omega_{2+}|
-
\frac 1 4
\Big[| \Omega_{2-}\rangle \langle \Omega_{2-}|,\rho\Big]_+
\right\}
+
\gamma_6
\left\{
\frac 1 2
|\Omega_-\rangle \langle \Omega_{2+}| \rho |\Omega_{2+}\rangle \langle \Omega_-|
-
\frac 1 4
\Big[| \Omega_{2+}\rangle \langle \Omega_{2+}|,\rho\Big]_+
\right\}
\nonumber
\\
&\pp=&%6
+
\gamma_7
\left\{
\frac 1 2
|\Omega_+\rangle \langle \Omega_{2-}| \rho |\Omega_{2-}\rangle \langle \Omega_+|
-
\frac 1 4
\Big[| \Omega_{2-}\rangle \langle \Omega_{2-}|,\rho\Big]_+
\right\}
+
\gamma_8
\left\{
\frac 1 2
|\Omega_-\rangle \langle \Omega_{2-}| \rho |\Omega_{2-}\rangle \langle \Omega_-|
-
\frac 1 4
\Big[| \Omega_{2-}\rangle \langle \Omega_{2-}|,\rho\Big]_+
\right\}.
\nonumber\\
\label{eq:rmaster}
\end{eqnarray}
\end{widetext}
Expressing the initial condition $\rho(0)=\sum_j c_j\rho_j$ as a linear combination of eigenvectors of $\cal L$, ${\cal L}\rho_j=\Lambda_j\rho_j$, we find the solution $\rho(t)=\sum_j c_je^{\Lambda_j t}\rho_j$. We need 25 eigenvectors. Twenty of them are easy to find and determine off-diagonal matrix elements (in the dressed-state basis),
\be
\rho_6
&=&%1
|\Omega_{2+}\rangle \langle \Omega_{2-}|,
\\
\Lambda_6
&=&
-i (\Omega_{2+} - \Omega_{2-})
-
\frac{ \gamma_4 + \gamma_5 + \gamma_6 + \gamma_7 + \gamma_8 + \gamma_e  }{4}, \nonumber
\\
\rho_7
&=&%2
|\Omega_{2+}\rangle \langle \Omega_+|,
\\
\Lambda_7
&=&-i (\Omega_{2+} - \Omega_+)
-
\frac{ \gamma_1 + \gamma_3 + \gamma_4 +\gamma_5 + \gamma_6 }{ 4 }
\ee
\be
\rho_8
&=&%3
|\Omega_{2+}\rangle \langle \Omega_-|,
\\
\Lambda_8
&=&
-i (\Omega_{2+} - \Omega_-)
-
\frac{ \gamma_c + \gamma_2 + \gamma_4 +\gamma_5 + \gamma_6  }{ 4 },\nonumber
\\
\rho_9
&=&%4
|\Omega_{2+}\rangle \langle \Omega_0|,
\\
\Lambda_9
&=&
-i (\Omega_{2+} - \Omega_0)
-
\frac{ \gamma_a + \gamma_b + \gamma_4 +\gamma_5 + \gamma_6  }{ 4 },\nonumber
\\
\rho_{10}
&=&%5
|\Omega_{2-}\rangle \langle \Omega_0|,
\\
\Lambda_{10}
&=&
-i (\Omega_{2-} - \Omega_0)
-
\frac{ \gamma_a + \gamma_b + \gamma_7 + \gamma_8  + \gamma_e  }{ 4 },\nonumber
\\
\rho_{11}
&=&%6
|\Omega_{2-}\rangle\langle \Omega_-|,
\\
\Lambda_{11}
&=&
-i (\Omega_{2-} - \Omega_-)
-
\frac{ \gamma_2 + \gamma_7 + \gamma_8 + \gamma_c + \gamma_e }{ 4 },\nonumber
\\
\rho_{12}
&=&%7
|\Omega_{2-}\rangle \langle \Omega_+|,
\\
\Lambda_{12}
&=&
-i(\Omega_{2-} - \Omega_+ )
-
\frac{ \gamma_1 + \gamma_3 + \gamma_7 + \gamma_8 + \gamma_e }{ 4 },\nonumber
\\
\rho_{13}
&=& %11
|\Omega_{2-}\rangle \langle \Omega_{2+}|,
\\\nonumber
\Lambda_{13}
&=&
i(\Omega_{2+} - \Omega_{2-} )
-
\frac{ \gamma_4 + \gamma_5 + \gamma_6 + \gamma_7 + \gamma_8 + \gamma_e  }{ 4 },
\\
\rho_{14}
&=&%12
|\Omega_+\rangle \langle \Omega_{2+}|,
\\\nonumber
\Lambda_{14}
&=&
i (\Omega_{2+} - \Omega_+)
-
\frac{ \gamma_1 + \gamma_3 + \gamma_4 +\gamma_5 + \gamma_6  }{ 4 },\nonumber
\\
\rho_{15}
&=&%13
|\Omega_+\rangle \langle \Omega_{2-}|,
\\\nonumber
\Lambda_{15}
&=&
i(\Omega_{2-} - \Omega_+ )
-
\frac{ \gamma_1 + \gamma_3 + \gamma_7 + \gamma_8 + \gamma_e }{ 4 },
\ee
\be
\rho_{16}
&=&%8
|\Omega_+\rangle \langle \Omega_-|,
\\
\Lambda_{16}
&=&
-i(\Omega_+ - \Omega_- )
-
\frac{ \gamma_1 + \gamma_2 + \gamma_3 + \gamma_c  }{ 4 },\nonumber
\\
\rho_{17}
&=&%9
|\Omega_+\rangle \langle \Omega_0|,
\\
\Lambda_{17}
&=&
-i(\Omega_+ - \Omega_0 )
-
\frac{ \gamma_1 + \gamma_3 + \gamma_a + \gamma_b  }{ 4 },\nonumber
\\\rho_{18}
&=&%10
|\Omega_-\rangle \langle \Omega_0|,
\\
\Lambda_{18}
&=&
-i(\Omega_- - \Omega_0 )
-
\frac{ \gamma_2 + \gamma_a +  \gamma_b + \gamma_c  }{ 4 },\nonumber
\\
\rho_{19}
&=&%14
|\Omega_-\rangle \langle \Omega_+|,
\\\nonumber
\Lambda_{19}
&=&
i(\Omega_+ - \Omega_- )
-
\frac{ \gamma_1 + \gamma_2 + \gamma_3 + \gamma_c   }{ 4 },
\\
\rho_{20}
&=&%15
|\Omega_-\rangle \langle \Omega_{2-}|,
\\\nonumber
\Lambda_{20}
&=&
i (\Omega_{2-} - \Omega_-)
-
\frac{ \gamma_2 + \gamma_7 + \gamma_8 + \gamma_c + \gamma_e }{ 4 },
\\
\rho_{21}
&=&%16
|\Omega_-\rangle \langle \Omega_{2+}|,
\\\nonumber
\Lambda_{21}
&=&
i (\Omega_{2+} - \Omega_-)
-
\frac{ \gamma_c + \gamma_2 + \gamma_4 +\gamma_5 + \gamma_6  }{ 4 },
\\
\rho_{22}
&=&%17
|\Omega_0\rangle \langle \Omega_{2+}|,
\\\nonumber
\Lambda_{22}
&=&
i (\Omega_{2+} - \Omega_0)
-
\frac{ \gamma_a + \gamma_b + \gamma_4 +\gamma_5 + \gamma_6  }{ 4 },
\\
\rho_{23}
&=&%18
|\Omega_0\rangle \langle \Omega_{2-}|,
\\\nonumber
\Lambda_{23}
&=&
i (\Omega_{2-} - \Omega_0)
-
\frac{ \gamma_a + \gamma_b + \gamma_7 + \gamma_8  + \gamma_e  }{ 4 },
\\
\rho_{24}
&=&%19
|\Omega_0\rangle \langle \Omega_+|,
\\\nonumber
\Lambda_{24}
&=&
i(\Omega_+ - \Omega_0 )
-
\frac{ \gamma_1 + \gamma_3 + \gamma_a + \gamma_b  }{ 4 },
\\
\rho_{25}
&=&%20
|\Omega_0\rangle \langle \Omega_-|,
\\\nonumber
\Lambda_{25}
&=&
i(\Omega_- - \Omega_0 )
-
\frac{ \gamma_2 + \gamma_a +  \gamma_b + \gamma_c  }{ 4 }.
\ee
The remaining five eigenvectors $\rho_j$, $j=1,2,3,4,5$ are related to the diagonal matrix elements of $\rho$,
\begin{eqnarray}
\rho_j
&=&%1
x_j |\Omega_{2+}\rangle \langle \Omega_{2+}|
+
y_j |\Omega_{2-}\rangle \langle \Omega_{2-}|
+
z_j |\Omega_+\rangle \langle \Omega_+|
\nonumber\\
&\pp=&
+
v_j |\Omega_-\rangle \langle \Omega_-|
+
w_j |\Omega_0\rangle \langle \Omega_0|
\label{eq:basis1}
\end{eqnarray}
The corresponding eigenvalue problem is equivalent to
\begin{widetext}
\begin{eqnarray}
\frac 1 2
\left(
\begin{array}{ccccc}
\gamma_4 + \gamma_5 + \gamma_6  & \gamma_e & 0 & 0 & 0\\
\gamma_5 & \gamma_7 + \gamma_8 + \gamma_e  & 0 &  0 & 0\\
\gamma_4 & \gamma_7 &  \gamma_1 + \gamma_3  & \gamma_c & \gamma_a\\
\gamma_6 & \gamma_8 & \gamma_3 &  \gamma_2 + \gamma_c & \gamma_b\\
0 & 0 & \gamma_1 & \gamma_2 &  \gamma_a + \gamma_b
\end{array}
\right)
\left(
\begin{array}{ccccc}
x_j\\
y_j\\
z_j\\
v_j\\
w_j
\end{array}
\right)
&=&
\Lambda_j
\left(
\begin{array}{ccccc}
x_j\\
y_j\\
z_j\\
v_j\\
w_j
\end{array}
\right).
\label{eq:macierz}
\end{eqnarray}
\end{widetext}
The eigenvalues are
\begin{subequations}
\begin{eqnarray}
\Lambda_1
&=&%1
0,
\\
\Lambda_2
&=&%2
-
0.25
\left(
\omega + \delta +\sqrt{\theta}
\right),
\\
\Lambda_3
&=&%3
-
0.25
\left(
\omega + \delta -\sqrt{\theta}
\right),
\\
\Lambda_4
&=&%4
-
0.25
\left(
\zeta + \xi +\sqrt{\kappa}
\right),
\\
\Lambda_5
&=&%5
-
0.25
\left(
\zeta + \xi -\sqrt{\kappa}
\right),
\end{eqnarray}
\end{subequations}
where
\begin{subequations}
\begin{eqnarray}
 \kappa
 &=&\gamma_6 ((\delta +\omega )^2-4 (\gamma_2(\gamma_3+\gamma_a){}
   \\
   \nonumber
   &+&(\gamma_a+\gamma_b) (\gamma_3+\gamma_c)+\gamma_1 (\gamma_2+\gamma_b+\gamma_c))),
   \end{eqnarray}
\begin{eqnarray}
    \theta=(\zeta -\xi )^2+4 \gamma_5 \gamma_e 
\end{eqnarray}
\begin{eqnarray}\omega=\gamma_1+\gamma_2+\gamma_3, \quad
\zeta=\gamma_4+\gamma_5+\gamma_6,
   \end{eqnarray}
\begin{eqnarray}\xi=\gamma_7+\gamma_8+\gamma_e, \quad
\delta=\gamma_a+\gamma_b+\gamma_c.
 \end{eqnarray}
 \end{subequations}

Explicit forms of the eigenvectors can be found in the Appendix.

A reasonable approximation of the initial thermal state is given by
\begin{eqnarray}
\rho(0)
&=&
0.95 |e,0\rangle \langle e,0| + 0.05 |e,1\rangle \langle e,1|\\
&=&%1
0.95
\Big(
 |\Omega_+\rangle \langle \Omega_+|
+
 |\Omega_-\rangle \langle \Omega_-|
\nonumber\\
&-&
 |\Omega_+\rangle \langle \Omega_-|
-
 |\Omega_-\rangle \langle \Omega_+|
\Big)
\nonumber\\
&+& 
0.05
\Big(
 |\Omega_{2+}\rangle \langle \Omega_{2+}|
+
 |\Omega_{2-}\rangle \langle \Omega_{2-}|
\nonumber\\
&-&
 |\Omega_{2+}\rangle \langle \Omega_{2-}|
-
 |\Omega_{2-}\rangle \langle \Omega_{2+}|
\Big).
\label{eq:3490}
\end{eqnarray}
After somewhat lengthy but simple calculations one finds that
\begin{eqnarray}
\rho(t)
&=&%1
2\big(B_1 e^{\Lambda_1 t} \rho_{1} + B_2 e^{\Lambda_2 t}  \rho_2 + B_3 e^{\Lambda_3 t}  \rho_3 + B_4 e^{\Lambda_4 t}  \rho_4
\nonumber\\
& & {}+ B_5 e^{\Lambda_5 t}  \rho_5 \big)+B_6 e^{\Lambda_6 t} \rho_6 +B_{13} e^{\Lambda_{13} t}  \rho_{13}\nonumber\\
& & {}
+ B_{16} e^{\Lambda_{16} t}  \rho_{16}+ B_{19} e^{\Lambda_{19} t}  \rho_{19}.
\label{rozwiazanie}
\end{eqnarray}
The coefficients are explicitly given in the Appendix. Probability $p_g(t)=p_{g,0}(t) + p_{g,1}(t) + p_{g,2}(t)$ of finding the atom in its ground state reads finally
\begin{eqnarray}
p_g(t)
&=&
 B_1 (x_{1} + y_1 + z_{1} + v_1 + 2 w_1) e^{\Lambda_1 t}\nonumber\\
& & {}
 +   B_2 (x_{2} + y_2 +z_{2} + v_2 + 2 w_2) e^{\Lambda_2 t}{}
\nonumber\\
& & {}+
  B_3 (x_{3} + y_3 + z_{3} + v_3 + 2 w_3) e^{\Lambda_3 t}\nonumber\\
& & {}+
 B_4 (x_{4} + y_4 +z_{4} + v_4 + 2 w_4)e^{\Lambda_4 t}{}
\nonumber\\
& & {}+
  B_5 (x_{5} + y_5 + z_{5} + v_5 + 2 w_5) e^{\Lambda_5 t}\nonumber\\
& & {}
-
0.025 e^{ - \frac {\gamma_4 + \gamma_5 +\gamma_6 +\gamma_7 +\gamma_8 +\gamma_e} {4} t} \cos{ 2 \sqrt{2} {\texttt{g}}  \text{t} } {}
\nonumber\\
& & {}-
0.475 e^{ - \frac {\gamma_1 + \gamma_2 +\gamma_3 +\gamma_c } {4} t} \cos{ 2 \sqrt{2} {\texttt{g}}  \text{t} }.
\label{eq:koniec}
\end{eqnarray}
Let us stress that the above solution is found under the assumption that the atom-field coupling $\texttt{g}$ is constant in time. We know, however, that the atom interacts with the mode whose spatial profile is Gaussian with width $w$. The atom is propagating through the cavity which makes, effectively, the coupling time-dependent. A method of taking this into account was discussed in detail in \cite{WC}. Assuming that the length of the cavity is $d$ we obtain probability appropriate for comparison with experimental data
\begin{eqnarray}
p_g(t)
&=&
 B_1 (x_{1} + y_1 + z_{1} + v_1 + 2 w_1) e^{\Lambda_1 t} \nonumber\\
& & {}+   B_2 (x_{2} + y_2 +z_{2} + v_2 + 2 w_2) e^{\Lambda_2 t}{}
\nonumber\\
& & {}+
  B_3 (x_{3} + y_3 + z_{3} + v_3 + 2 w_3) e^{\Lambda_3 t}\nonumber\\
& & {}+
  B_4 (x_{4} + y_4 +z_{4} + v_4 + 2 w_4)e^{\Lambda_4 t}{}
\nonumber\\
& & {}+
  B_5 (x_{5} + y_5 + z_{5} + v_5 + 2 w_5) e^{\Lambda_5 t}
\nonumber\\
& & {}-
0.025 e^{ - \frac {\gamma_4 + \gamma_5 +\gamma_6 +\gamma_7 +\gamma_8 +\gamma_e} {4} t} \cos
\big(
{ 2 \sqrt{2} {\texttt{g}} {\sqrt{\pi} \frac {w} {d}} \text{t} }
\big) {}
\nonumber\\
& & {}-
0.475 e^{ - \frac {\gamma_1 + \gamma_2 +\gamma_3 +\gamma_c  } {4} t} \cos
\big(
{ 2 \sqrt{2} {\texttt{g}} {\sqrt{\pi} \frac {w} {d}} \text{t} }
\big).
\label{eq:moj}
\end{eqnarray}
Let us recall that the data shown in \cite{Brune} were plotted as a function of an effective time $t_{\texttt{eff}}=\sqrt{\pi} \frac{w}{d} t$.
Since it is more convenient for us to work in terms of $t$ than $t_{\texttt{eff}}$ \cite{WC}, we rescale the data to $t$. Fig.~4 shows vacuum Rabi oscillation predicted by our more realistic scenario (solid line) as compared to experimental data and predictions of the simplified model from \cite{WC} (dotted).

\begin{center}
\begin{figure}[h]
		\includegraphics[width=8cm]{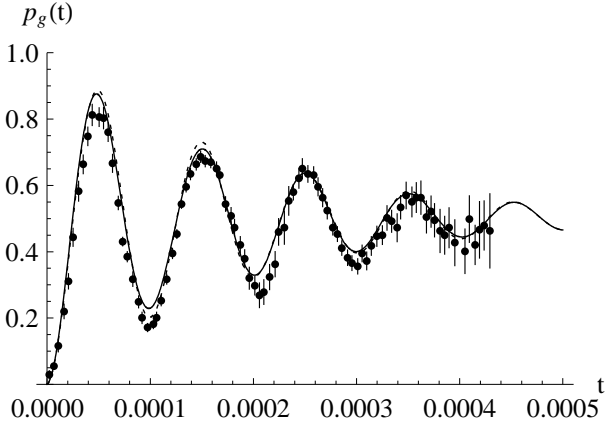}
	\caption{The solid line represents (\ref{eq:moj}). The dotted curve is the prediction from \cite{WC}. The parameters are $\gamma_1=\gamma_2=\gamma_4=\gamma_6=\gamma_7=\gamma_8 = 17.73$ Hz,$\gamma_a=\gamma_b=\epsilon \gamma_1$, $\gamma_3=\gamma_c=\gamma_5=\gamma_e = 0.07 \texttt{g}$, $\texttt{g}=47 \pi 10^3$ Hz. $t$ is the true time.}
	\label{fig:moj}
\end{figure}
\end{center}

\section{Conclusions}

The effort devoted to solving the more realistic case apparently did not pay: The new curve does not describe the data any better. The approximation made in \cite{WC} turns out to be physically reasonable. This is in a sense good news since inclusion of yet higher dressed doublets would require solving algebraic equations of order higher than 5, and thus there would be practically no chance for finding exact solutions.

On the other hand, this also means that refined theoretical analysis of Rabi oscillation has not brought us any closer to understanding why probability of atomic exited state does not decay to zero as fast as expected on the grounds of cavity lifetime reported in \cite{Brune}. The cavity quality seems to be better than that assumed by Brune {\it et al.\/} Perhaps, as suggested in \cite{WC}, the key element is in opening of the cavity and inclusion of the long-wave transitions within a given doublet of dressed states. The appropriate jump operators occur if one modifies the cavity-reservoir coupling. The required interaction with the reservoir has stronger dependence on the number of photons inside of the cavity. Measurements of cavity lifetimes are typically performed with more cavity photons than in Rabi oscillation experiments. It is possible that the same cavity has a much longer lifetime if only a few photons are present. The problem requires further experimental studies.

\section*{Appendix}
\noindent
Coefficients introduced in previous section \\are: ${B_6=B_{13}=-0.025}$ 
and ${B_{16}=B_{19}=-0.025}$,  \\ whereas for ${n=1,2,3,4,5}$ we have:
\be
B_n&=&\frac{1}{\chi}\left[ 0.2375(A_{3n}+A_{4n})+0.0125 (A_{1n}+A_{2n})\right]
\nonumber
\ee
where
\begin{eqnarray}
A_{1i}&=&\sum_{\stackrel{\{j,k,l,m\}=1}{\{j,k,l,m\}\neq i}}^5{\varepsilon_{jklm}\:y_j\:z_k\:v_l\:w_m} \quad   (i=1,..,5)
\nonumber
\ee
\be
A_{2j}&=&\sum_{\stackrel{\{i,k,l,m\}=1}{\{i,k,l,m\}\neq j}}^5{\varepsilon_{iklm}\:x_i\:z_k\:v_l\:w_m} \quad   (j=1,..,5)
\nonumber
\ee
\be
A_{3k}&=&\sum_{\stackrel{\{i,j,l,m\}=1}{\{i,j,l,m\}\neq k}}^5{\varepsilon_{ijlm}\:x_i\:y_j\:v_l\:w_m} \quad  (k=1,..,5)
\nonumber
\ee
\be
A_{4l}&=&\sum_{\stackrel{\{i,j,k,m\}=1}{\{i,j,k,m\}\neq l}}^5{\varepsilon_{ijkm}x_i\:y_j\:z_k\:w_m} \quad   (l=1,..,5)
\nonumber
\ee
\be
A_{5m}&=&\sum_{\stackrel{\{i,j,k,l\}=1}{\{i,j,k,l\}\neq m}}^5{\varepsilon_{ijkl}\:x_i\:y_j\:z_k\:v_l} \quad   (m=1,..,5)
\nonumber
\end{eqnarray}
\begin{eqnarray}
\chi=\sum_{\{i,j,k,l,m\}=1}^5{\varepsilon_{ijklm}\:x_i\:y_j\:z_k\:v_l\:w_m}
\nonumber
\end{eqnarray}
\noindent
For eigenvalues ${\Lambda_k}$ (${k=1,2,3,4,5}$) we define eigenvectors with coordinates ${\{x_k,y_k,z_k,v_k,w_k\}}$
\begin{subequations}
\begin{eqnarray}
x_1
&=&%1
0
\nonumber
\\
y_1
&=&%2
0
\nonumber
\\
z_1
&=&%3
\gamma_2 \gamma_a +\gamma_c \gamma_a + \gamma_b \gamma_c
\nonumber
\\
v_1
&=&%4
\gamma_3 \gamma_a +\gamma_1 \gamma_b + \gamma_3 \gamma_b
\nonumber
\\
w_1
&=&%5
\gamma_1 \gamma_2 +\gamma_2 \gamma_3 + \gamma_1 \gamma_c
\nonumber
\end{eqnarray}
\end{subequations}
\noindent
For ${k=2,3}$ we have
\begin{subequations}
\begin{eqnarray}
x_k
&=&
0
\nonumber
\\
y_k
&=&
0
\nonumber
\\
z_k
&=&
32\gamma_1\left(a_z+(-1)^k \sqrt{\kappa}\:b_z \right)
\nonumber
\\
v_k
&=&
32\gamma_1\left(a_v+(-1)^k \sqrt{\kappa}\:b_v \right)
\nonumber
\\
w_k&=&
32\gamma_1\left(a_w+(-1)^k \sqrt{\kappa}\:b_w \right)
\nonumber
\end{eqnarray}
\end{subequations}
and if ${k=4,5}$ we have
\begin{subequations}
\begin{eqnarray}
x_k
&=&
16\gamma_1\gamma_6\left(c_x +(-1)^k \sqrt{\theta}\:d_x \right)
\nonumber
\\
y_k
&=&
32\gamma_1 \gamma_5 \gamma_6\left(c_y +(-1)^k \sqrt{\theta}\:d_y \right)
\nonumber
\\
z_k
&=&
32\gamma_1\gamma_6\left(c_z +(-1)^k \sqrt{\theta}\:d_z \right)
\nonumber
\\
v_k
&=&
32\gamma_1 \gamma_6 \left(c_v +(-1)^k \sqrt{\theta}\:d_v \right)
\nonumber
\\
w_k&=&
128 \gamma_1 \gamma_6 \left(c_w +(-1)^k \sqrt{\theta}\:d_w \right)
\nonumber
\end{eqnarray}
\end{subequations}
where
 
\begin{subequations}
\begin{eqnarray}
a_z
&=&
2 \gamma_5 (\gamma_6 \gamma_7-\gamma_4 \gamma_8) \big(\omega ^2-\delta ^2-2 \gamma_2(\delta
   +\omega ) 
    {}
\nonumber
\\ 
&+&
4 (\gamma_2 \gamma_a+\gamma_c \gamma_a+\gamma_b \gamma_c)\big)+
 (\gamma_6(\delta+\omega -2 \xi ) {}
\nonumber
\\ 
&-&2 \gamma_5 \gamma_8)
\big(4 \gamma_6 \gamma_c^2+2 \gamma_6 \gamma_c(\omega -\delta ) -4 \gamma_2\gamma_a (\gamma_4+\gamma_6) {}
\nonumber
\\ 
&+&\gamma_4 \big(\delta ^2-\omega ^2
   \gamma_2+2 (\delta +\omega )-4\gamma_c (\gamma_a+\gamma_b) \big)\big){}
\nonumber
\\ 
&-& \kappa \gamma_6 (\gamma_4(\delta +3 \omega -2 \xi )-2
   (\gamma_2 \gamma_4+\gamma_5 \gamma_7+\gamma_6 \gamma_c)) 
   \nonumber
\\
   a_v
   &=&
   4 \gamma_5 (\gamma_4 \gamma_8-\gamma_6 \gamma_7) (\gamma_3(\omega -\delta ) +2
   (\gamma_3 \gamma_c-\gamma_1 \gamma_b)) {}
\nonumber
\\ 
&-&\kappa \gamma_6 (\gamma_6(\delta +\omega -2 \xi )-2 (\gamma_3
   \gamma_4+\gamma_5 \gamma_8{}
\nonumber
\\ 
&-&\gamma_6 (\gamma_2+\gamma_c)))+((\delta +\omega -2 \xi ) \gamma_6-2 \gamma_5 \gamma_8) {}
\nonumber
\\ 
&\times&(2 \gamma_4
   (\gamma_3(\omega -\delta )+2 (\gamma_3 \gamma_c-\gamma_1 \gamma_b)){}
\nonumber
\\ 
&+&\gamma_6 ((\gamma_1 +\gamma_3 -\gamma_a-\gamma_b)^2-(\gamma_2+\gamma_c)^2+4 \gamma_1 \gamma_a))
\nonumber
\\ 
   a_w
   &=&
   2\big((\gamma_6(\delta +\omega -2 \xi ) -2 \gamma_5 \gamma_8) (\gamma_1 \gamma_4(\delta +\omega )  {}
\nonumber
\\ 
&+& \gamma_2 \gamma_6(\delta -\omega )-2 \gamma_2 (\gamma_1 \gamma_4+\gamma_3 \gamma_4-\gamma_2 \gamma_6){}
\nonumber
\\ 
&-&2 \gamma_1 \gamma_c(\gamma_4+\gamma_6) )+2 \gamma_5 (\gamma_6
   \gamma_7-\gamma_4 \gamma_8){}
\nonumber
\\ 
&\times& (2 (\gamma_1 \gamma_2+\gamma_3 \gamma_2+\gamma_1
   \gamma_c)-\gamma_1(\delta +\omega ) ){}
\nonumber
\\ 
&+&\kappa \gamma_6 (\gamma_1 \gamma_4+\gamma_2 \gamma_6) \big)
   \nonumber
\\
  b_z
  &=&
  4  \gamma_6 ((\gamma_1+\gamma_3) (\gamma_4(\xi -\omega ) +\gamma_2 \gamma_4+\gamma_5 \gamma_7){}
\nonumber
\\ 
&-&\gamma_a(\gamma_1 \gamma_4 +\gamma_2 \gamma_6)+(\omega -\xi +\gamma_c) \gamma_6
   \gamma_c{}
\nonumber
\\ 
&-&\gamma_c (\gamma_3 \gamma_4+\gamma_5 \gamma_8))
\nonumber
\\ 
   b_v
   &=&
   4  \gamma_6 (\gamma_3 \gamma_4 (\omega -\xi +\gamma_c)-\gamma_b(\gamma_1 \gamma_4+\gamma_2 \gamma_6){}
\nonumber
\\ 
&+&(\gamma_5 \gamma_8+\gamma_6 (\xi -\gamma_2-\gamma_c)) (\gamma_2+\gamma_c) {}
\nonumber
\\ 
&-&\gamma_3 (\gamma_5 \gamma_7+\gamma_6 \gamma_c))
\nonumber
\\ 
    b_w
    &=&
    4  \gamma_6 (\gamma_2 \gamma_6( \delta-\xi ) -\gamma_2 (\gamma_1 \gamma_4+\gamma_3
   \gamma_4-\gamma_2 \gamma_6{}
\nonumber
\\ 
&+&\gamma_5 \gamma_8)-\gamma_1 (\gamma_4 (\xi-\delta -\omega 
   +\gamma_c)- \gamma_5 \gamma_7-\gamma_6 \gamma_c))
\nonumber
\\
c_x
   &=&
  \eta  \left(\zeta ^2-\xi ^2\right) +\theta (\theta+\eta-\xi ^2 {}
\nonumber
\\ 
&+&\zeta ( 4 \xi-4 \delta -4 \omega +5 \zeta )) 
\nonumber
  \\
   c_y
   &=&\theta  (2 (\delta +\omega )-3 (\xi +\zeta ))-\eta  (\xi +\zeta ) 
   \nonumber
\end{eqnarray}
\end{subequations}

\begin{subequations}
\begin{eqnarray}
   c_z
   &=&
   4 \gamma_2  ((\zeta -\xi ) (\gamma_4 \gamma_b-\gamma_6 \gamma_a)+2 \gamma_5 (\gamma_8 \gamma_a-\gamma_7 \gamma_b)) {}
\nonumber
\\ 
&+& ( \xi +\zeta-2 \delta +2 \gamma_c) ((2 \gamma_5 \gamma_7-\gamma_4(\zeta-\xi ) ){}
\nonumber
\\ 
&\times&(\xi +\zeta -2 \gamma_2-2 \gamma_c) +2\gamma_c (\gamma_6(\zeta -\xi ) -2 \gamma_5 \gamma_8) ){}
\nonumber
\\ 
&-&\theta   (\gamma_4( \xi +3 \zeta-2 \delta ) -2 (\gamma_2 \gamma_4+\gamma_5 \gamma_7+\gamma_6 \gamma_c))
\nonumber
\\
c_v&=&-\theta  (\gamma_6 (-2 (\delta +\omega -\gamma_2-\gamma_c)+3 \zeta +\xi )- 2
   (\gamma_3 \gamma_4  {}
\nonumber
\\ 
&+&\gamma_5 \gamma_8))+(-2 \delta +\zeta+\xi +2 \gamma_c)(2 \gamma_3
   ((\zeta -\xi )
   {}
\nonumber
\\ 
&\times&
    (\gamma_4+\gamma_6)+2 \gamma_5 (\gamma_e-\xi ))-(\zeta +\xi-2 \gamma_1){}
\nonumber
\\ 
&\times&
   (\gamma_6 (\zeta -\xi )-2 \gamma_5 \gamma_8))-4 \gamma_1 ((\zeta -\xi ){}
\nonumber
\\ 
&\times& (\gamma_4\gamma_b-\gamma_6 \gamma_a)+2 \gamma_5 (\gamma_8 \gamma_a-\gamma_7 \gamma_b)) \nonumber
\\
 c_w  &=& \gamma_4+\gamma_2 \gamma_6)-(\gamma_1 \gamma_2+\gamma_1 \gamma_c+\gamma_2 \gamma_3)((\zeta -\xi ) {}
\nonumber
\\
&\times&(\gamma_4+\gamma_6)- 2 \gamma_5 \gamma_8)+\gamma_5
   \gamma_7 (2 \gamma_2 (\gamma_1+\gamma_3) {}
\nonumber
\\ 
&-&-\gamma_1 (\zeta +\xi -2 \gamma_c))-\gamma_2 \gamma_5 \gamma_8 (\zeta +\xi )
\nonumber
\\   d_x
   &=&2 ((\xi +\zeta-\delta -\omega  )\left(\zeta ^2-\xi ^2\right){}
\nonumber
\\ 
&-&\theta  (\delta +\omega-\xi  -2 \zeta )+\eta  \zeta ) 
\nonumber
\\
   d_y
   &=&
    2 (\delta +\omega -\xi -\zeta ) (\xi +\zeta )-\eta -\theta 
   \nonumber
\\  
 d_z
   &=&
   4 (\gamma_5 \gamma_7( \xi +\zeta-\delta -\gamma_2) +\zeta  ( \gamma_4(\delta -\zeta +\gamma_2)+\gamma_6 \gamma_c){}
\nonumber
\\ 
&-&
 \gamma_5(\gamma_8 \gamma_c+\gamma_4 \gamma_e)
 - (\gamma_2 \gamma_a+\gamma_c(\gamma_a+\gamma_b) )(\gamma_4+\gamma_6)) 
\nonumber
\\
  d_v
  &=&
   4 (-\gamma_5 (\gamma_3 \gamma_7-\gamma_2 \gamma_8+\gamma_6 \gamma_e)-(\gamma_4+\gamma_6)(\gamma_3 \gamma_a{}
\nonumber
\\ 
&+& \gamma_b(\gamma_1+\gamma_3))+\gamma_5
   \gamma_8 ( \xi +\zeta -\delta-\omega +\gamma_c){}
\nonumber
\\ 
&+&\zeta  (\gamma_3 \gamma_4+\gamma_6 (\delta +\omega -\zeta -\gamma_2-\gamma_c)))
\nonumber
\\
   d_w
   &=&
    \gamma_2 \gamma_6(\zeta -\omega ) -\gamma_2 (\gamma_3 \gamma_4-\gamma_2 \gamma_6+\gamma_5 \gamma_8){}
\nonumber
\\ 
&+&\gamma_1 \gamma_4 (\zeta -\gamma_2-\gamma_c)-\gamma_1
   (\gamma_5 \gamma_7+\gamma_6 \gamma_c)
   \nonumber
 \\
\eta
&=&
\gamma_2 (4 \gamma_3-2 (\xi +\zeta -2 \gamma_a)){}
\nonumber
\\
&+&(\xi +\zeta -2 \gamma_3-2 \gamma_c) ( \xi +\zeta-2 \delta +2 \gamma_c){}
\nonumber
\\
&+&2\gamma_1 (2 \gamma_2- ( \xi +\zeta -2 \delta+2 \gamma_a))
\nonumber
\end{eqnarray}
\end{subequations}

\end{document}